\documentclass[aps,pra,twocolumn,superscriptaddress,showpacs,floatfix]{revtex4}
\usepackage[dvipdfmx]{graphicx,color}
\begin{document}

\title{Compensation of gravity on cold atoms by a linear optical potential}
\author{Kosuke Shibata}
\email{shibata@qo.phys.gakushuin.ac.jp}
\affiliation{Department of Physics, Gakushuin University Tokyo, Japan}
\author{Hidehiko Ikeda}
\affiliation{Department of Physics, Gakushuin University Tokyo, Japan}
\author{Ryota Suzuki}
\affiliation{Department of Physics, Gakushuin University Tokyo, Japan}
\author{Takuya Hirano}
\affiliation{Department of Physics, Gakushuin University Tokyo, Japan}

\date{\today}

\begin{abstract}
We demonstrate gravity compensation for an ultracold gas of $^{87}$Rb atoms
with a time-averaged optical potential.
The position of a far-off-resonance beam is temporally modulated with an acousto-optic deflector 
to efficiently produce a potential with a linear gradient independent of the atomic magnetic sublevels. 
We realize compensation of the gravity sag and 
preparation of a degenerate gas in a trap with weak vertical confinement.
Optical gravity compensation will provide the opportunity to perform experiments under microgravity in a laboratory
and broaden the scope of cold atom research.
\end{abstract}

\maketitle
\section{Introduction}
Cold atoms are significantly affected by gravity due to their low kinetic energy.
While cold atoms can be used to measure fundamental constants such as the Newtonian gravitational constant \cite{Rosi2014} and 
offer interesting subjects such as the gravity on a quantum gas,
gravity often induces undesired effects in experiments.
This limits the size, shape and density of a trapped atom cloud 
because of the requirement to vertically hold atoms against gravity.
It also deteriorates the performance of precision measurements 
using atoms, e.g., in atomic interferometry \cite{D'Amico2017}.

A microgravity environment, free of these limitations, offers a rich ground for science and applications.
Microgravity environments have been realized
in ballistic flight \cite{Stern2009}, tall drop towers \cite{Zoest2010,Muntinga2013}, 
and space \cite{Becker2018}.
A magnetic field gradient is a convenient tool
for gravity cancellation on atoms in a magnetically sensitive atomic state.
Magnetic gravity cancellation has been used for producing dilute Bose--Einstein condensates (BEC) below 500 pK \cite{Leanhardt2003}, a slow atom laser \cite{Buning2010} and a homogeneous 3D gas \cite{Gaunt2013}.

The production of microgravity environments, regardless of the atomic states,
will broaden the scope of experiments in laboratories.
As an important application, 
dilute or homogeneous gases with spin degrees of freedom can be produced.
A homogeneous spinor gas is a good platform for precise magnetometry
due to the absence of density-dependent inhomogeneous collisional shifts \cite{Harber2002}.
It is also useful for a precise matter-wave interferometer \cite{Cronin2009}.

The optical potential due to a far-detuned light field is state-independent \cite{GRIMM200095}
and should be a promising component for convenient and versatile gravity cancellation.
Optical gravity cancellation to achieve a microgravity environment has not been realized, however.
A difficulty for optical gravity cancellation is to create a sufficiently linear intensity gradient 
over an atom cloud (typically $10^{2}$ to $10^{3}$ $\mu$m).
With a Gaussian beam,
gravity cancellation for an entire gas requires a beam of very large size and an impractically enormous power, 
although the reduction of a gravity sag has been implemented \cite{Konishi2016}.
Spatial light modulators, which have recently been widely used in cold atom experiments 
to produce optical traps with various configurations \cite{Gaunt2013,Nogrette2014,Gauthier2016},
can in principle create a linear intensity profile.
However, it might be a challenge to 
achieve a beam profile using a spatial light modulator of good linearity over a wide area covering an atom cloud.

Here, we propose gravity cancellation by a time-averaged linear optical potential, produced by deflecting a beam with an acousto-optic deflector (AOD) driven by a frequency-modulated radio-frequency (rf) wave.
Such a ``painting'' technique \cite{Henderson2009} has been used for the
construction of integrated coherent matter wave circuits \cite{Ryu2015} and 
the production of dynamically shaped atom traps for rapid evaporative cooling \cite{Roy2016}.
As we show below, 
a beam having a linear intensity gradient suitable for gravity compensation can be produced with the painting technique.
We apply a beam with a linear intensity gradient to compensate the gravity sag of an ultracold $^{87}$Rb atom gas in an optical trap.
Furthermore, we prepare a degenerate gas in a trap with weak vertical confinement with the aid of the compensation beam.

The paper is organized as follows.
We explain the method for producing an intensity profile with a linear gradient by the painting technique in Sec.~\ref{sec: method}. 
In Sec.~\ref{sec: result}, we describe the experimental results for
the produced intensity profiles and the gravity compensation on cold atom clouds.
The potential applications are discussed in Sec.~\ref{sec: discussion}.
We conclude the paper in Sec.~\ref{sec: conclusion}.

\section{Method}\label{sec: method}

\begin{figure}
 \centering
 \includegraphics[width=8cm]{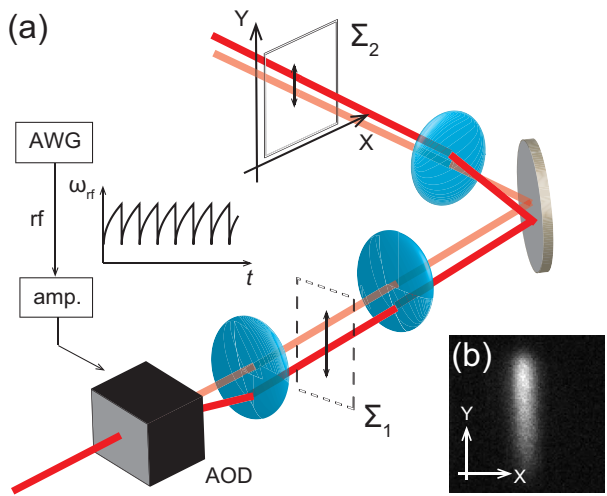}
 \caption{(Color online) (a) Setup for optical painting.
A frequency modulated rf wave from an arbitrary waveform generator (AWG)
is sent to the AOD. 
The beam deflected by the AOD is focused by a lens on the plane $\Sigma_1$.
The deflected beam displacement at $\Sigma_1$ is proportional to the change in the rf frequency.
The beam is demagnified onto the target plane $\Sigma_2$ in a vacuum glass cell (not shown) using a lens pair.
 (b) An example 2D intensity profile of a painted beam. 
The field of view is $200 \times 200$ $\mu$m.
}
\label{fig: op}
\end{figure}

Before presenting the experimental details, 
we describe how to design a target intensity profile. 
The scheme relies on the simple fact that
the slower (faster) the beam crosses a certain position, 
the larger (smaller) the light intensity at the position.
We design the intensity profile through frequency modulation (FM) of the rf wave
applied to the AOD.
We only modulate the rf frequency, to efficiently use the beam power,
whereas amplitude modulation allows more flexible generation of a desired profile.

Let us consider the time-averaged intensity produced by 
a beam scanned along the $y$-direction.
We first consider the case in which the beam size is sufficiently small.
Its position at time $t$ is designated by $y(t)$
and the intensity distribution at $t$ is given by $I(y,t) = I_0\delta(y-y(t))$.
In this case, the intensity averaged from $t=0$ to $t=T$, $\bar{I}(y) \equiv \frac{1}{T} \int_0^T I(y,t) dt$, 
is inversely proportional to $|\dot{y}(t)|$.
The trajectory $y(t)$ giving a target averaged intensity $\bar{I}(y)$ is in general given by the solution of
\begin{equation}
\frac{C}{\dot{y}(t)}= \bar{I}(y)
\end{equation}
with $C$ being a constant.
Therefore, a linearly increasing intensity is produced with a trajectory satisfying $y(t) - y(0) \propto \sqrt{t}$.
When $y(t)$ responds linearly to the rf frequency (which holds for a small beam displacement), the FM waveform $\nu(t)$ to produce a linear intensity profile is given by
\begin{equation}\label{eq: nu}
\nu(t) = b \sqrt{t} + \nu(0),
\end{equation}
where $b$ is a constant.
 
We can produce a profile with a linear slope by frequency modulation using Eq. (\ref{eq: nu}) with a Gaussian beam with a finite beam waist $w$.
The averaged intensity of the beam with its center scanned 
from $y(0)$ to $y(T) = y(0) +L$
according to Eq. (\ref{eq: nu}) is written as
\begin{eqnarray}\label{eq: I}
\frac{\bar{I}(y)}{I_0} &=& \sqrt{\pi} \beta \eta \left[ \mathrm{erf} \left( \frac{1-\eta}{\beta} \right) +  \mathrm{erf} \left( \frac{\eta}{\beta} \right)   \right] \nonumber\\
&& +\beta^2 \left[ e^{-\eta^2/\beta^2} - e^{-(1-\eta)^2/\beta^2} \right],
\end{eqnarray}
where $I_0$ is the peak beam intensity without modulation,
$\eta=\frac{y}{L}$, $\beta = \frac{w}{\sqrt{2}L}$ and $\mathrm{erf} (z) = \frac{2}{\sqrt{\pi}}\int_0^z e^{-t^2} dt$.
As $\beta$ becomes smaller, better linearity is achieved.
In other words, a linear intensity gradient is produced 
when the scan width $L$ is sufficiently larger than $w$.
With a scan width of $L=7w$, 
the deviation from a perfect linear intensity 
is smaller than $10^{-6}$ over a range of 3 $w$.
When $L=20w$, the relative deviation can be smaller than 
$10^{-14}$ over a range of 13 $w$.

The experimental setup for the optical painting is shown in Fig.~\ref{fig: op}(a).
The compensation beam source is a fiber Bragg grating (FBG) stabilized laser 
with a center wavelength of $\lambda = 976$ nm (Thorlabs, BL976-PAG900),
producing an attractive potential for Rb atoms.
The beam is deflected by a two-axis AOD (IntraAction Corp., DTD-274HD6M).
An rf waveform with its frequency modulated according to Eq. (\ref{eq: nu}) 
is repeatedly generated by an arbitrary waveform generator (Lecroy, ArbStudio 1102)
and sent to the AOD through a mixer for light power adjustment and an rf amplifier.
The deflected beam is focused by a lens of focal length 100 mm
after the AOD, and is demagnified onto the target plane in a vacuum glass cell
($\Sigma_2$ in Fig.~\ref{fig: op}(a)) through a pair of lens with focal lengths of 500 mm and 200 mm.

\section{Experimental Results}\label{sec: result}

\begin{figure} 
 \centering
 \includegraphics[width=8.5cm]{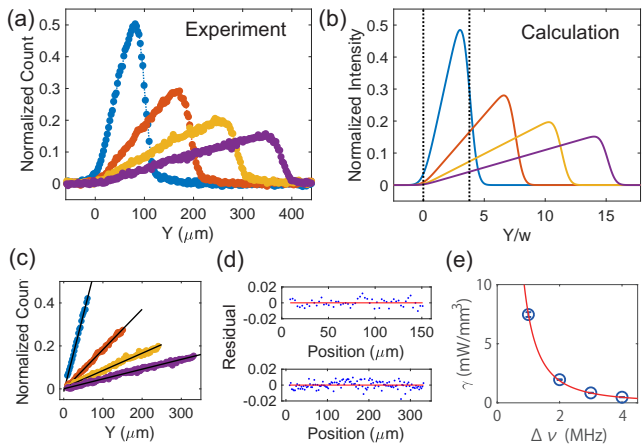}
 \caption{(Color online) Horizontally integrated intensity of the painted beam.
(a) Experimental results. Data with frequency scan widths of 1 MHz, 2 MHz, 3 MHz and 4 MHz are represented by the blue, orange, yellow and purple points, respectively.
                 The intensity is normalized by the peak intensity without modulation.
              (b) Calculated intensity. The calculation for the scan widths 
                   $(3.5, 7, 10.5, 14) w$ are represented by the blue, orange, yellow and purple lines, respectively,
                   where $w$ is the beam waist and measured to be 24.7(7) $\mu$m.
               The vertical dotted lines indicate the boundaries of the scanned beam center for a scan width of 3.5$w$. 
              (c) Linear fits to the slope regions of measured profiles.
              (d) Residuals of the linear fits for scan widths of 2 MHz [upper panel] and 4 MHz [lower panel].
              (e) Relation between the intensity gradient $\gamma$ 
                  and the frequency scan width $\Delta\nu$.
                   The red solid curve is a fit using $a\Delta \nu^{-2}$ (see the text).
}
\label{fig: profile}
\end{figure}

We examine the time-averaged intensity of the scanned beam 
to confirm the validity of the method and 
to evaluate the performance of the experimental system.
We take images of the beam using an imaging system consisting of lenses and a charge-coupled device (CCD) camera after the cell.
The camera exposure time is set to 20 ms, sufficiently longer than the frequency scan cycle of 100 $\mu$s, to measure the effective average intensity profile.
A typical measured 2D intensity profile with $\Delta \nu$ = 2 MHz is shown in Fig.~\ref{fig: op}(b).
The vertical intensity distributions obtained by integrating the 2D profile along the horizontal ($X$) axis are plotted in Fig.~\ref{fig: profile}(a). 
The observed profiles with frequency scan widths of $\Delta \nu =$ 1 MHz, 2 MHz, 3 MHz and 4 MHz are in good agreement with the calculated profiles
based on Eq. (\ref{eq: I}) shown in Fig.~\ref{fig: profile}(b).
 
The slope regions are fitted by a linear function to determine the gradient (see Fig.~\ref{fig: profile}(c)).
We observe no significant systematic deviation from the linear fit, as shown in Fig.~\ref{fig: profile}(d), 
implying good linearity of the beam intensity. 
The fitting residuals mainly come from noise in the imaging.
The intensity gradient per unit power, $\gamma$, is plotted as a function of the scan width in Fig.~\ref{fig: profile}(e).
We fit the gradient by $a \Delta\nu^{-2}$, the validity of which we confirm with the calculated profiles,
obtaining $a$ = 7.49(16) mW mm$^{-3}$ (MHz)$^{-2}$.
It is estimated from this result that
a 980-nm beam with $P$ = 173 mW with $\Delta\nu$ = 1.5 MHz should cancel gravity on $^{87}$Rb atoms.

\begin{figure}
 \centering
 \includegraphics[width=8.5cm]{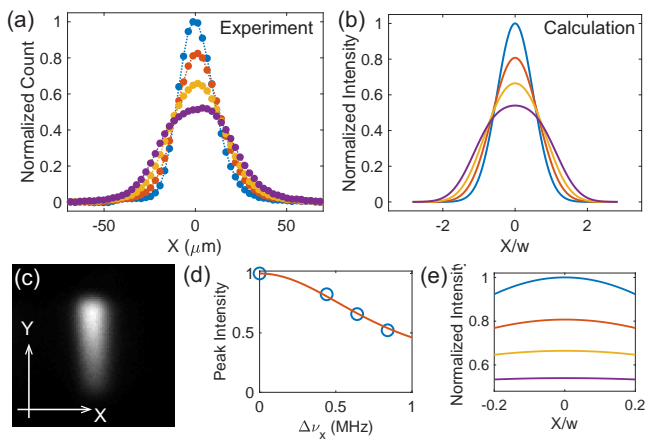}
 \caption{(Color online) Vertically integrated intensity of the painted beam.
             (a) Experimental results. The vertically integrated intensities with $\Delta\nu_x =$ 0 MHz, 0.44 MHz, 0.64 MHz and 0.84 MHz are represented by the blue, orange, yellow and purple points, respectively.
               (b) Calculated intensity. The calculations for horizontal scan widths $(0, 1.23, 1.79, 2.35) w$
                 are represented by the black, blue, green and red solid lines, respectively.
               (c) 2D intensity profile with $\Delta\nu_x = 0.64$ MHz. The field of view is $200 \times 200$ $\mu$m.
               (d) Measured peak intensity (blue circles)
                    and calculated peak intensity (red solid line) as a function of $\Delta \nu_x$.
               (e) Magnification of the calculated intensity around $X=0$.  
}
\label{fig: xscan}
\end{figure}

The above intensity profiles are actually not very suitable for compensating gravity on a cold atom cloud,
because the potential is \textit{horizontally} inhomogeneous.
As a countermeasure,
we sweep the rf frequency on the horizontal AOD axis linearly in addition to the vertical modulation to realize a top-flat horizontal beam profile.
To be specific, we apply horizontal and vertical frequency modulations given by
\begin{equation}\label{eq: nux}
\nu_x(t) = \nu_{x0} + \Delta\nu_x \frac{t}{T_x}, 
\end{equation}
\begin{equation}\label{eq: nuy}
\nu_y(t) = \nu_{y0} + \Delta\nu_y \sqrt{\frac{t}{T_y}},
\end{equation}
respectively.
We set $T_x $= 8 $\mu$s and $T_y$ = 250 $\mu$s to satisfy
$T_x \ll T_y$, keeping the repetition rates $T_x^{-1}$ and $T_y^{-1}$ much higher than the trap frequencies.

Fig.~\ref{fig: xscan}(c) shows the intensity profile for frequency modulation
with $\Delta\nu_x = 640$ kHz and $\Delta\nu_y = 1.5$ MHz.
Vertically integrated intensity distributions with different $\Delta\nu_x$ are shown in Fig.~\ref{fig: xscan}(a).
The corresponding calculated profiles are given by 
\begin{equation}
\tilde{I}(X) = \frac{\sqrt{\pi}\tilde{I}(0)} {4 \alpha} 
                \left[ \mathrm{erf}\left(\chi + \alpha \right) -  \mathrm{erf}\left(\chi - \alpha \right) \right]
\end{equation}
where $\chi =\sqrt{2} X/w$ and $\alpha= L_X/(\sqrt{8}w)$ 
with $L_X$ being the horizontal scan width,
and are plotted in Fig.~\ref{fig: xscan}(b), showing good agreement with the experimental results.
When $\Delta\nu_x = 0.64$ MHz, the intensity inhomogeneity in $|X| <$ 3 $\mu$m is reduced to less than $10^{-2}$ (see Fig.~\ref{fig: xscan}(e)),
at the cost of a decrease of the peak intensity by $1/3$, as shown in Fig.~\ref{fig: xscan}(d).

We apply the painted beam to a cold atom cloud in a crossed optical trap \cite{Shibata2019}.
The atoms are transferred from a magnetic trap to the optical trap
and are cooled by forced evaporative cooling to degeneracy.
We prepare a partially Bose-condensed gas of $3 \times 10^5$ atoms in the $\vert F, m \rangle =\vert 2, 2 \rangle$ state.
The trap frequencies of the crossed optical trap 
are $(\omega_x,\omega_y,\omega_z) = 2\pi \times$ (340, 70, 30) Hz. 
After the evaporation, we ramp up the compensation beam power, $P$, during a period of 100 ms,
and hold the atoms for another 100 ms.
Then, we release the atoms from the trap 
and take an absorption image after a time-of-flight (TOF).
The beam power is stabilized and the power fluctuation is below $3 \times 10^{-3}$.

\begin{figure}
 \centering
 \includegraphics[width=8.5cm]{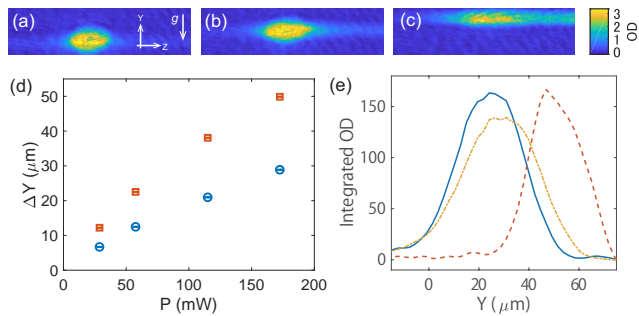}
 \caption{(Color online) Absorption images of the atom clouds after a TOF of 0.2 ms
(a) with almost no compensation, (b) with a compensation beam with 
$P$ = 116 mW and $\Delta\nu_y =$1.5 MHz and (c) with a beam with $P$= 174 mW and $\Delta\nu_y$ = 1 MHz. 
The field of view in each panel is 100 $\mu$m $\times$ 400 $\mu$m.
(d) Displacement of the vertical cloud center. 
The data with $\Delta\nu_y$ = 1 MHz and 1.5 MHz
are represented by the red circles and blue squares, respectively.
The solid blue line is a linear fit to the data with $\Delta \nu_y$ = 1.5 MHz.
(e) Horizontally integrated optical densities of (b) [blue solid line] 
and (c) [red dashed line] 
The yellow dashed-dotted line represents the density for a weakened trap.
 }
\label{fig: dY}
\end{figure}

We first present results for scan widths of $\Delta\nu_x$ = 0.64 MHz and $\Delta\nu_y$ = 1.5 MHz.
The atoms move upward by the application of the compensation beam, as can be seen 
by comparing the cases with $P=$ 2 mW (Fig.~\ref{fig: dY}(a))
and 116 mW (Fig.~\ref{fig: dY}(b)).
We plot the vertical displacement of the cloud center, $\Delta Y$, in Fig.~\ref{fig: dY}(d) [blue circles].
The position for the case with $P=$ 2 mW is taken as the origin of the displacement. 
While the displacement is smaller than or comparable to the vertical waist size of the trap beam ($\sim$ 70 $\mu$m), 
the original potential can be approximated by a parabola
and the linear compensation beam should lead to a displacement proportional to $P$.
The observed displacement is consistent with the fact that gravity sag is compensated by the linear optical potential.

We also apply a beam with $\Delta\nu_y = 1$ MHz, which produces a steeper intensity gradient.
The displacement is larger than the case with  $\Delta\nu_y = 1.5$ MHz, as expected, but tends to saturate for higher $P$.
This is because the compensation is so great that the atoms are pulled up to the upper boundary of the painted potential. 
In fact, when a beam with $P=174$ mW is applied,
the vertical width of the atom cloud becomes narrow, as can be seen in Figs.~\ref{fig: dY}(c) and \ref{fig: dY}(e),
resulting from the combined potential of gravity and the painted beam.
We also observe for higher $P$ that 
atoms leak from the crossed region due to potential distortion in the side region
(see Fig.~\ref{fig: dY}(c)).
This is further evidence that the beam power is sufficient for gravity compensation, 
because the vertical confinement in the side region is provided by a single axial optical trap
and cannot trap atoms without gravity compensation.

If gravity is properly canceled, 
the atoms can be held in a trap with weak vertical confinement.
We demonstrate the expansion of atoms into a weaker confinement
with the aid of a compensation beam.
In this experiment,
we set the scan widths to $\Delta \nu_x$ = 0.8 MHz and $\Delta \nu_y$ = 1.3 MHz.
During the ramp up of the compensation beam over  300 ms,
we decrease the axial and beam power from 131 mW to 29 mW 
and the radial beam power from 58 mW to 38 mW.
The mean trap frequency of the weakened trap is estimated 
from the total number of atoms and the critical temperature
to be 2$\pi \times$ 59 Hz,  $\sim$2/3 of the original mean trap frequency.

\begin{figure} 
 \centering
 \includegraphics[width=8.5cm]{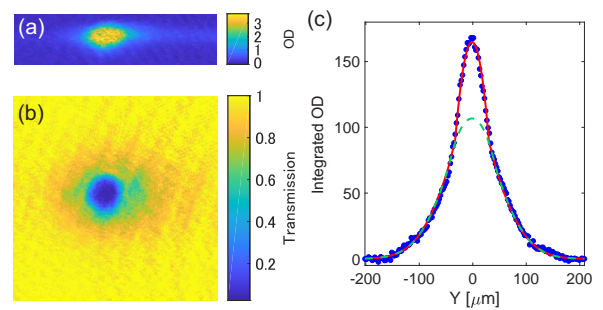}
 \caption{(Color online) Absorption images with TOFs of  (a) 0.2 ms and (b) 16 ms.
  (c) Horizontally integrated optical density of the data shown in (b). 
      The red solid and green dashed lines are a fitting curve using a bimodal function and its thermal component, respectively. 
 }
\label{fig: bimodal}
\end{figure}

While the crossed optical trap with decreased power alone cannot hold atoms
because of the severe potential distortion due to gravity,
we observe that the atoms remain after a hold time of 600 ms 
in the weakened optical trap with the compensation beam.
An image obtained with a TOF of 0.2 ms and $P=200$ mW is shown in Fig.~\ref{fig: bimodal}(a).
The vertical cloud size becomes slightly larger
(see Fig.~\ref{fig: dY}(e)), reflecting the weaker vertical confinement.
The gas is a partially degenerate BEC.
By a bimodal fitting to the image with a TOF of 16 ms (see Figs.~\ref{fig: bimodal}(b) and \ref{fig: bimodal}(c)),
we obtain a total number of atoms of 2.8 $\times 10^5$ and a condensate fraction of 0.29. 
The measured temperature is $T$ = 142 $\pm$ 7 nK. 
We can further decrease the trap beam power
to produce a BEC with no discernible thermal component.
The persistence of the BEC indicates that the painted linear potential does not cause severe heating of the atoms \cite{Onofrio2000}.

\section{Discussion}\label{sec: discussion}
The demonstrated optical gravity compensation will broaden the scope of cold atom research.
Cold atom gases can be prepared in weak or homogeneous traps regardless of the spin.
This opens the possibility of preparing gases of various spin components
in weak or homogeneous traps, including homogeneous spinor gases.
Gravity compensation for spinless alkaline-earth(-like) atoms can also be realized.
The method can be applied to construct compact precise interferometers
requiring a long-time expansion of atoms.
We estimate the volume in which gravity is compensated.
A laser with a power of 3 W and a wavelength of 808 nm should compensate gravity on Rb atoms over a volume of  
0.5 mm $\times$ 0.5 mm $\times$ 0.5 mm,
which will enable free expansion over several hundreds of milliseconds, 
comparable to that in experiments in tall towers \cite{Zoest2010,Muntinga2013} 
and space \cite{Becker2018}.
For precise measurements, the use of a compensation beam at a magic wavelength would be required
and the gravity compensation volume may be reduced due to the limited beam power.
The limitation, however, might be relaxed with improvements in laser technology.

\section{Conclusion}\label{sec: conclusion}
We demonstrate compensation of gravity on cold atoms 
with a time-averaged linear optical potential.
The optical gravity cancellation will open new possibilities for cold atom experiments,
including the production of very dilute gases, the study of homogeneous gases with spin degrees of freedom, and precise measurements.

\begin{acknowledgements}
This work was supported by the Quantum Leap Flagship Program (MEXT Q-LEAP) and
JSPS KAKENHI Grant Number JP18K45678.
We would like to thank Munekazu Horikoshi for his inspiring comments on the applications of this work.
\end{acknowledgements}


\end{document}